\documentclass[prl,twocolumn,showpacs,floatfix, superscriptaddress]{revtex4}

\usepackage[figuresright]{rotating}
\usepackage{psfig, amssymb, graphicx, epsfig, rotate, subfigure, dcolumn, bm}

\begin{document}

\newcommand*{\atow}[0]{$\alpha$$\to$$\omega$}
\newcommand*{\eps}[0]{\varepsilon}
\newcommand*{\NEBVASP}[0]{\textit{ab~initio} NEB}

\title{A New Mechanism for the $\alpha$ to $\omega$ Martensitic
Transformation in Pure Titanium}

\author{D. R. Trinkle}
\affiliation{Ohio State University, Columbus, OH 43210}
\author{R. G. Hennig}
\affiliation{Ohio State University, Columbus, OH 43210}
\author{S. G. Srinivasan}
\affiliation{Los Alamos National Laboratory, Los Alamos, NM 87545}
\author{D. M. Hatch}
\affiliation{Brigham Young University, Provo, UT 84602}
\author{M. D. Jones}
\affiliation{State University of New York, Buffalo, NY 14260}
\author{H. T. Stokes}
\affiliation{Brigham Young University, Provo, UT 84602}
\author{R. C. Albers}
\affiliation{Los Alamos National Laboratory, Los Alamos, NM 87545}
\author{J. W. Wilkins}
\affiliation{Ohio State University, Columbus, OH 43210}

\date{\today}

\begin{abstract}
We propose a new direct mechanism for the pressure driven \atow\ martensitic
transformation in pure titanium. A systematic algorithm enumerates all
possible mechanisms whose energy barriers are evaluated.
A new, homogeneous mechanism emerges with a barrier at least four
times lower than other mechanisms.
This mechanism remains favorable in a simple nucleation model.
\end{abstract}


\maketitle

Martensitic transformations are abundant in nature and are commonly
used in engineering technologies~\cite{martensites}.
Materials from steel to shape-memory alloys are governed by their
underlying martensitic transformations~\cite{shape-memory}.
The pressure driven $\alpha\text{(hcp)}\to 
\omega\text{(hexagonal)}$ transformation in pure titanium, discussed
here and reviewed extensively by Sikka 
\textit{et~al.}~\cite{sikka}, has significant technological
implications in the aerospace industry because the $\omega$ phase
formation lowers toughness and ductility. This transformation was
first observed by Jamieson~\cite{jamieson}, and has since been
extensively studied using static high-pressure~\cite{static} and
shockwave methods~\cite{shockwave}.  Because of experimental
difficulties in directly observing martensitic transformation
mechanisms, they are usually inferred from the orientation
relationships between the initial and final phases. Such an approach
may result in multiple transformation mechanisms for any given set of
orientation relations, and requires one to guess the appropriate
transformation mechanism.  Thus, despite several attempts, the
mechanism for this transformation is still unclear.

We calculate the energy barrier for homogeneous transformation for
different titanium \atow\ transformation mechanisms and compare the
values using a simplified nucleation model.  We systematically
generate and sort possible \atow\ mechanisms by their energy barriers.
A new direct mechanism emerges whose barrier is lowest both
homogeneously and when considered in a simple nucleation model.

\begin{figure*}[htb]
\includegraphics[width=7.0in]{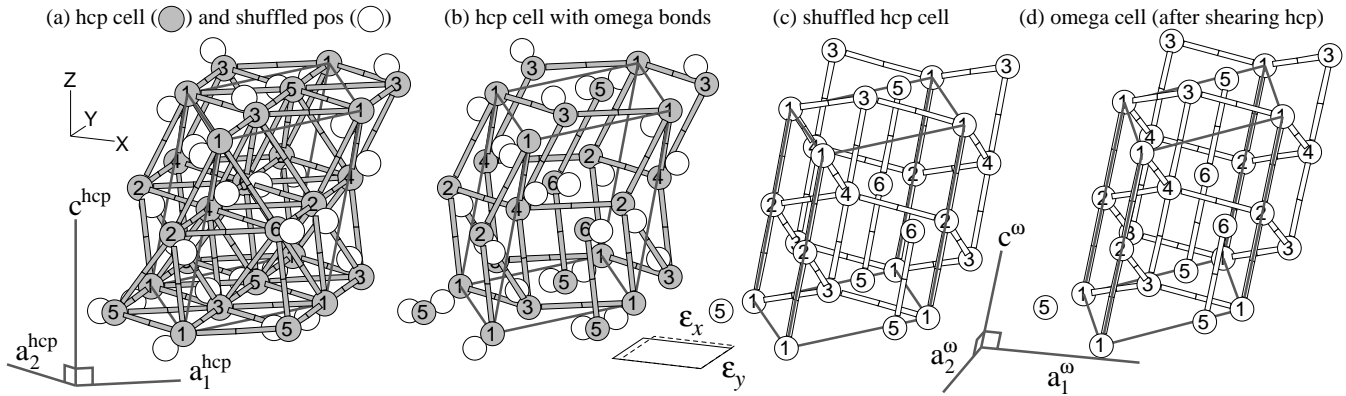}
\caption{
Our proposed \atow ~transformation mechanism (TAO-1).  The direct
6-atom transformation is visualized (i) as a two-step process, (ii)
with 21 additional atoms outside the heavy gray parallelpiped
supercell.  In $\alpha$, atoms 1,3,5 and 2,4,6 are in the A and B
stacking planes, respectively; whereas in $\omega$, atoms 1--4 are in
the Wyckoff $d$ position of space group $P6/mmm$, and atoms 5--6 are
in the Wyckoff $a$ position~\cite{spacegroup}.  (a) The grey atoms in
the $\alpha$ cell shuffle to new positions (white atoms), with atoms
1--4 shuffling 0.63~\AA, and atoms 5--6 shuffling 0.42~\AA.  (b) The
$\alpha$ cell is redrawn with the ``bonds'' of the $\omega$ structure.
\textbf{Step 1:} Shuffling the grey atoms to the white positions, the
$\alpha$ cell (b) produces a strained $\omega$ cell (c), contained in the same
supercell.
\textbf{Step 2:} Straining the supercell (c)
$\eps_x=0.91$, $\eps_y=1.12$, and $\eps_z=0.98$ produces the final
$\omega$ supercell (d).  The orientation relations connecting the $\alpha$ and
$\omega$ supercells are: $(0001)_\alpha \;\|\; (0\bar 111)_\omega$ 
and $[11\bar 20]_\alpha \;\|\; [01\bar 11]_\omega$.}
\label{Tao1}
\end{figure*}

Figure~\ref{Tao1} shows our new low energy barrier mechanism for the
\atow\ transformation in Ti, called TAO-1, for ``Titanium Alpha to
Omega.''  This direct 6-atom transformation requires no intermediary
phase, and has small shuffles and strains.  The 6 atoms divide into a
group of 4 atoms that shuffle by 0.63~\AA\ and 2 atoms that shuffle by
0.42~\AA.  Combining these shuffles with strains of $\eps_x=0.91$,
$\eps_y=1.12$, and $\eps_z=0.98$ produces a final $\omega$ cell from
our $\alpha$ cell.  The original $\alpha$ matrix is then oriented
relative to the $\omega$ matrix such that $(0001)_\alpha \;\|\; (0\bar
111)_\omega$ and $[11\bar 20]_\alpha \;\|\; [01\bar 11]_\omega$.
These orientation relations are seen in some experiments, but not
others~\cite{static,shockwave,usikov}.

Our mechanism identification method matches possible supercells of
$\alpha$ and $\omega$ to determine the lattice strain, and atom
positions to determine the necessary internal relaxations; similar
to~\cite{comsubs}.  While there are infinitely many possible
supercells, we consider only 6 and 12 atom supercells with principal
strains less than 1.333 and greater than $0.75 = 1.333^{-1}$.  For
each supercell, there are multiple ways to shuffle the atom positions
from $\alpha$ to $\omega$; we consider mechanisms where, relative to
the center of mass, no atom moves more than 2.0~\AA.  We also limit
the closest distance of approach between any two atoms during the
transformation.  Each supercell has one mechanism with the largest
closest approach distance; we consider all mechanisms for that
supercell within 0.1~\AA\ of that value.

To efficiently compare the enthalpy barriers of the possible mechanisms,
we use three methods of increasing computational
sophistication and accuracy. 
\textbf{1.}
For each mechanism, we calculate an approximate barrier based only on
the strain, using the elastic approximation.  Such a calculation of
the ``elastic barrier'' relies only on the elastic constants for each
phase, and the supercell geometry%
~\footnote{If $E_{\mathrm{s-}\alpha}$ and $E_{\mathrm{s-}\omega}$ are
the energies of the fully strained $\alpha$ and $\omega$ within the
elastic approximation, the elastic barrier is $E_{\mathrm{s-}\alpha}
E_{\mathrm{s-}\omega} / (E_{\mathrm{s-}\alpha}^{{1/3}} +
E_{\mathrm{s-}\omega}^{{1/3}})^3$.}.
\textbf{2.}
For mechanisms with a low elastic barrier, we calculate an energy
landscape using a strain variable ($\eps$) and a single concerted
shuffle variable ($\delta$).  The energies $E(\eps, \delta)$ are
calculated using a tight-binding (TB) model~\cite{tight-binding} on a
grid between $\alpha$ at $E(0,0)$ and $\omega$ at $E(1,1)$.  The
energy at the transition state in this reduced space gives a
``landscape barrier.''
\textbf{3.}
For mechanisms with the lowest landscape barriers, we calculate the
exact barrier using the nudged elastic band (NEB)
method~\cite{jonsson} together with our TB model~\cite{tight-binding}
or \textit{ab~initio} VASP simulations~\cite{VASP}, or both.  During
the NEB calculation, the cell shape and size are allowed to change in
response to the applied pressure.  In general, we expect the elastic
barrier to underestimate the TB landscape barrier, and the TB
landscape barrier to overestimate the TB-NEB barrier.  The most
probable mechanism should have the smallest
\NEBVASP\ enthalpy barrier.

We calculate total energies using a TB model and carefully converged
\textit{ab~initio} calculations.  The TB calculations are performed
with OHMMS~\cite{ohmms} and use Mehl and Papaconstantopoulos's
functional form~\cite{tight-binding} with parameters that are refit to
reproduce LAPW total energies for hcp, bcc, fcc, omega, and sc to
within 0.5~meV/atom.  For each mechanism, we use a k-point mesh
equivalant to $12\times12\times8$ in hcp with a 63~meV thermal
smearing to give an energy convergence of 1~meV/atom.  For our
\textit{ab~initio} calculations~%
\footnote{VASP~\cite{VASP} is a plane-wave based code using ultra-soft
Vanderbilt type pseudopotentials~\cite{pseudo1} as supplied by
G. Kresse and J. Hafner~\cite{pseudo2}.  The calculations were
performed using the generalized gradient approximation of Perdew and
Wang~\cite{perdew}.}
we include $3p$ electrons in the valence band, and use a plane-wave
kinetic-energy cutoff of 400~eV and a $7\times 7\times 7$ k-point mesh
to ensure energy convergence to within 1~meV/atom.  We relax the
atomic positions and the unit cell shape and volume until the total
electronic energy changes by less than 1~meV.

\begin{figure}
\includegraphics[height=2.0in]{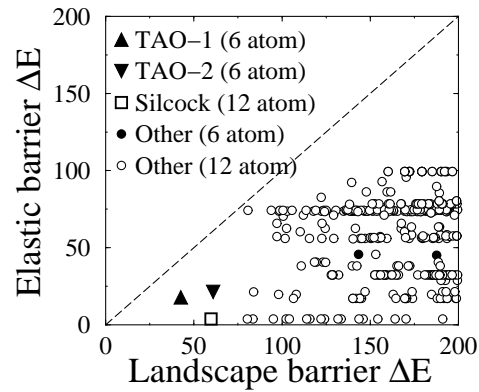}
\caption{Approximate elastic energy barrier (meV/atom) versus 
landscape energy barrier (meV/atom) for 359 mechanisms whose landscape
barrier is below 200 meV/atom.  Of the 134 supercells generated, 57
had low (less than 100 meV/atom) elastic barriers.  From those 57
supercells, 977 mechanisms were generated; for those 977 mechanisms,
we calculate a shear/shuffle energy landscape using our tight-binding
parameters.  Of those 977, 359 mechanisms had an elastic barrier below
200 meV/atom.  We select the three mechanisms in the bottom left
corner---TAO-1, TAO-2, and Silcock---and calculate their true barrier
using the \NEBVASP\ method.}
\label{figEnComparison}
\end{figure}

The elastic and landscape energy barrier calculations reduce our
initial set of 977 candidate mechanisms to the 3 lowest barrier mechanisms.
From the initial list of 134 supercells, we reject all but 57 because
their elastic barriers are large (greater than 100~meV/atom).  For the
remaining 57 supercells, we generate 977 mechanisms (6 6-atom
mechanisms and 971 12-atom mechanisms), and compute TB landscape
barriers.
Figure~\ref{figEnComparison} shows that the elastic barrier
underestimates the landscape barrier.
We select the three mechanisms with the smallest landscape
barriers---TAO-1, TAO-2, and Silcock---for careful study.

\begin{figure}
\subfigure[\label{figSilcockHCP} $\alpha$ phase]
  {\includegraphics[width=1.65in]{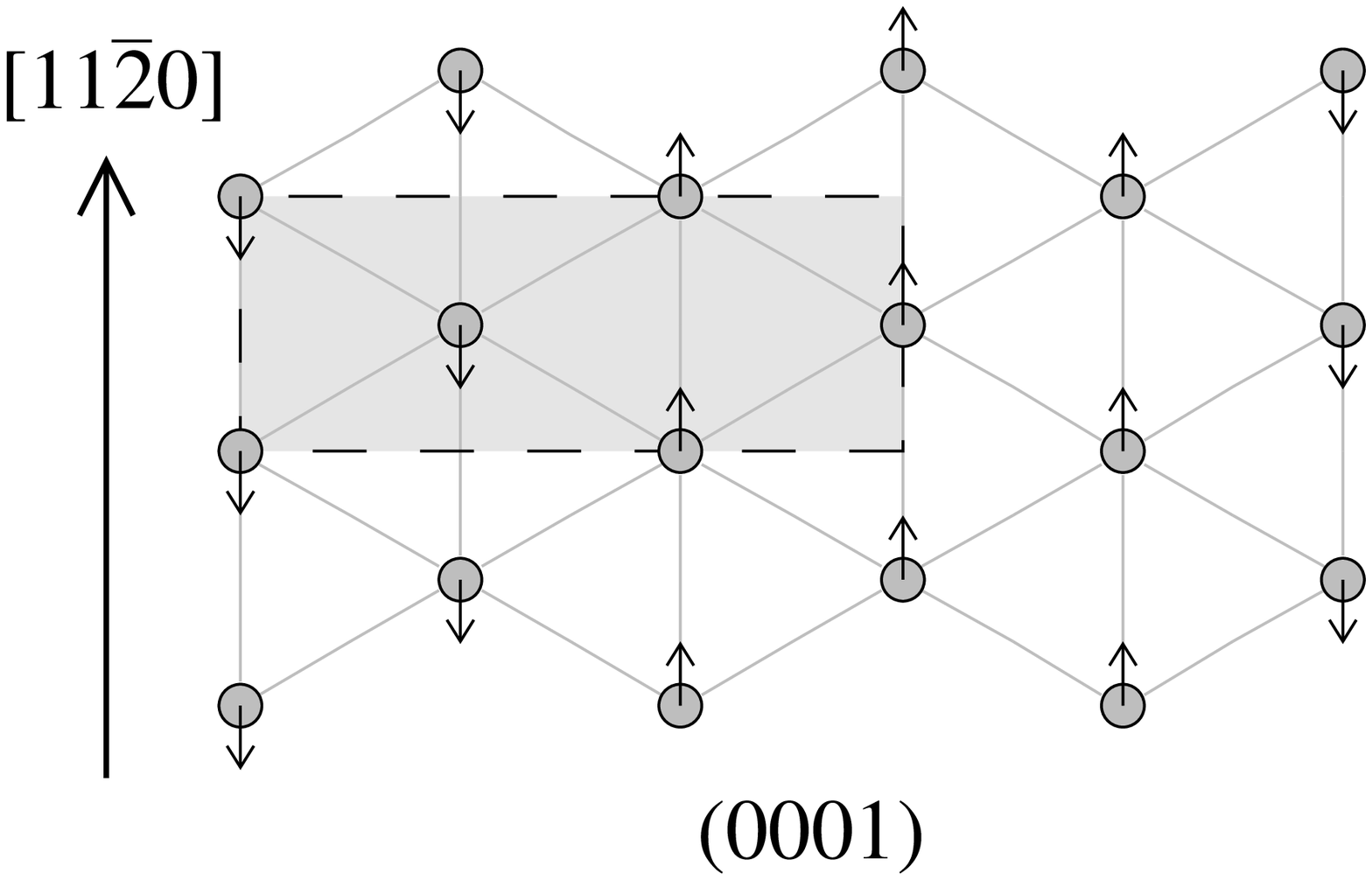}}
\subfigure[\label{figSilcockOmega} $\omega$ phase]
  {\includegraphics[width=1.65in]{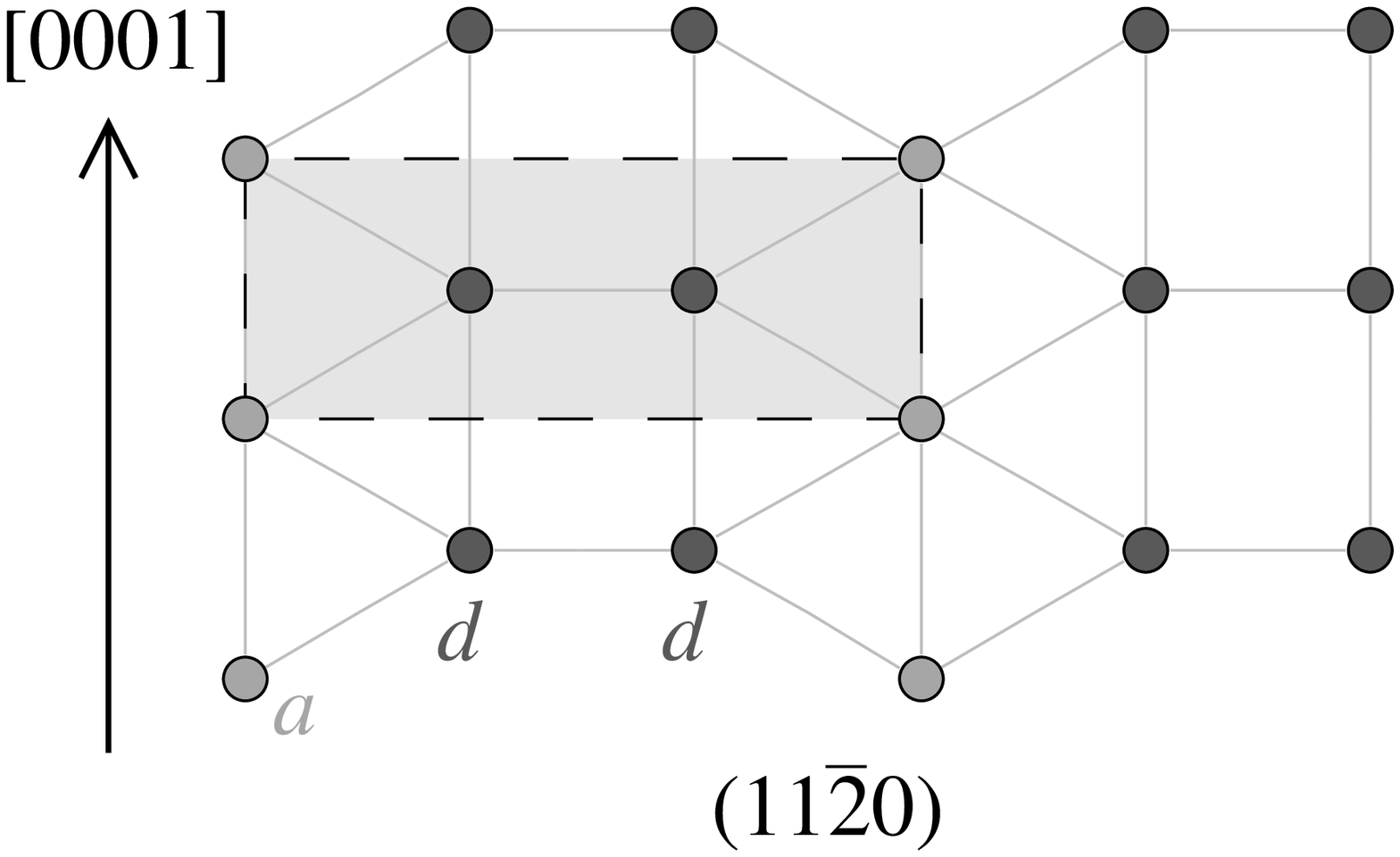}}
\caption{\label{figSilcockHCPtoOmega}
Silcock mechanism for $\alpha$ to $\omega$ transformation (single
$\alpha$ stacking plane shown).  (a) In each stacking plane, 3 out of
every 6 atoms shuffle by 0.74~\AA\ along $[11\bar 20]_\alpha$, while
the other 3 shuffle in the opposite direction $[\bar 1\bar
120]_\alpha$.  (b) This creates the $(11\bar 20)_\omega$ plane from
$(0001)_\alpha$.  The $a$ and $d$ Wyckoff positions of $P6/mmm$ are
also shown.  This shuffle is accompanied by a strain $\eps_x=1.05$
along $[1\bar 100]_\alpha$ and $\eps_y=0.95$ along $[11\bar
20]_\alpha$ to produce a hexagonal $\omega$ cell with the correct
$c/a$ ratio.  The orientation relations connecting the $\alpha$ and
$\omega$ supercells are: $(0001)_\alpha \;\|\; (11\bar 20)_\omega$ and
$[11\bar 20]_\alpha \;\|\; [0001]_\omega$.}
\end{figure}

Figure~\ref{figSilcockHCPtoOmega} illustrates the Silcock
mechanism~\cite{silcock}.  Her mechanism, determined from observed
orientation relations, involves significant atomic shuffle, relatively
small strains, and is a direct transformation mechanism with no
intermediate state. In each $\alpha$ stacking plane, 3 out of every 6
atoms shuffle by 0.74~\AA\ along $[11\bar 20]_\alpha$, while the other
3 shuffle in the opposite direction $[\bar 1\bar 120]_\alpha$.  This
shuffle is accompanied by a strain $\eps_x=1.05$ along $[1\bar
100]_\alpha$ and $\eps_y=0.95$ along $[11\bar 20]_\alpha$ to produce a
hexagonal $\omega$ cell with the correct $c/a$ ratio.  Despite the
lack of direct conclusive evidence to support the Silcock mechanism,
it is still widely invoked to describe the \atow\ transformation
because it is a direct mechanism~\cite{silcock-evidence}.

The remaining two mechanisms, TAO-1 and TAO-2, are related to Usikov
and Zilbershtein's~\cite{usikov} proposed composite mechanism
involving the thermodynamically unstable $\beta\text{(bcc)}$
intermediary phase, and proceeding as
$\alpha$$\to$$\beta$$\to$$\omega$.  Using the $\alpha$-$\omega$
orientation relations, they proposed an $\alpha$$\to$$\beta$
transformation via an inverse Burgers mechanism~\cite{burgers}
followed by a $\beta$$\to$$\omega$ transformation via the collapse of
2 out of 3 $(111)_\beta$ planes.  This produced two unique mechanisms,
called Variant~I and Variant~II, depending on the direction of
$\lbrace 111\rbrace_\beta$ planes to collapse.  The TAO-1 and TAO-2
mechanisms could have been constructed from Variant I and II,
respectively, by using the $\beta$ intermediate system to construct a
6 atom supercell.  Allowing Variant I and II to relax away from the
$\beta$ phase would have resulted in direct transformation mechanisms.

Table~\ref{table:mechEn} summarizes the energy barriers for the three
mechanisms of interest and identifies TAO-1 as having the lowest
energy barrier.  The TAO-1 mechanism provides the best trade off
between shuffle and strain; during the transformation, the closest
nearest neighbor distance is 2.63~\AA, which is larger than the
2.55~\AA\ value for TAO-2 and 2.57~\AA\ for Silcock.  Because the
nearest-neighbor distances in $\alpha$ and $\omega$ are 2.95~\AA\ and
2.65~\AA, respectively, it is not surprising that TAO-1 has the lowest
barrier.  The barrier to pass through the intermediate $\beta$ phase,
as suggested by Usikov and Zilbershtein~\cite{usikov}, is
108~meV/atom---drastically larger than TAO-1's barrier of 9~meV/atom.

\begin{table}
\caption{Comparison of TAO-1, TAO-2, and Silcock mechanisms.
Energy barriers: Four different methods for calculating
the energy barrier for the three mechanisms are shown, from least
accurate to most accurate.  The elastic barrier only accounts for the
strain in each mechanism.  The landscape barrier uses a simple
combined shuffle for each, and a tight-binding total energy.  Finally,
the NEB calculation is done with the tight-binding method and
\textit{ab~initio} to accurately determine the barrier. 
Orientation Relations: The relative orientation of $\alpha$ to
$\omega$ is shown for each mechanism.  N.B.: Silcock and TAO-2
mechanisms have the same orientation relations.}
\label{table:mechEn}
    \begin{tabular}{lc@{\quad}cc}
      \toprule
                                      & TAO-1   & TAO-2   & Silcock \\
      \colrule
      \multicolumn{4}{c}{}\hfill Homogeneous barriers (in meV/atom)\hfill\\
      Elastic barrier:       & 18  & 21  &  3.7  \\
      Landscape barrier:     & 41  & 59  & 59    \\
      TB-NEB barrier:        & 24  & 52  & 54    \\
      \NEBVASP\ barrier:     &  9  & 58  & 31    \\
      \colrule
      \multicolumn{4}{l}{}\hfill Transformation information\hfill\\
      Orientation      & $(0001)_\alpha \;\|\; (0\bar 111)_\omega$ &
      \multicolumn{2}{c}{$(0001)_\alpha \;\|\; (11\bar 20)_\omega$} \\
      \quad Relations: & $[11\bar 20]_\alpha  \;\|\; [01\bar 11]_\omega$ &
      \multicolumn{2}{c}{$[11\bar 20]_\alpha  \;\|\; [0001]_\omega$} \\
      \botrule
    \end{tabular}
\end{table}

\begin{figure}[tb]
\includegraphics[width=3.25in]{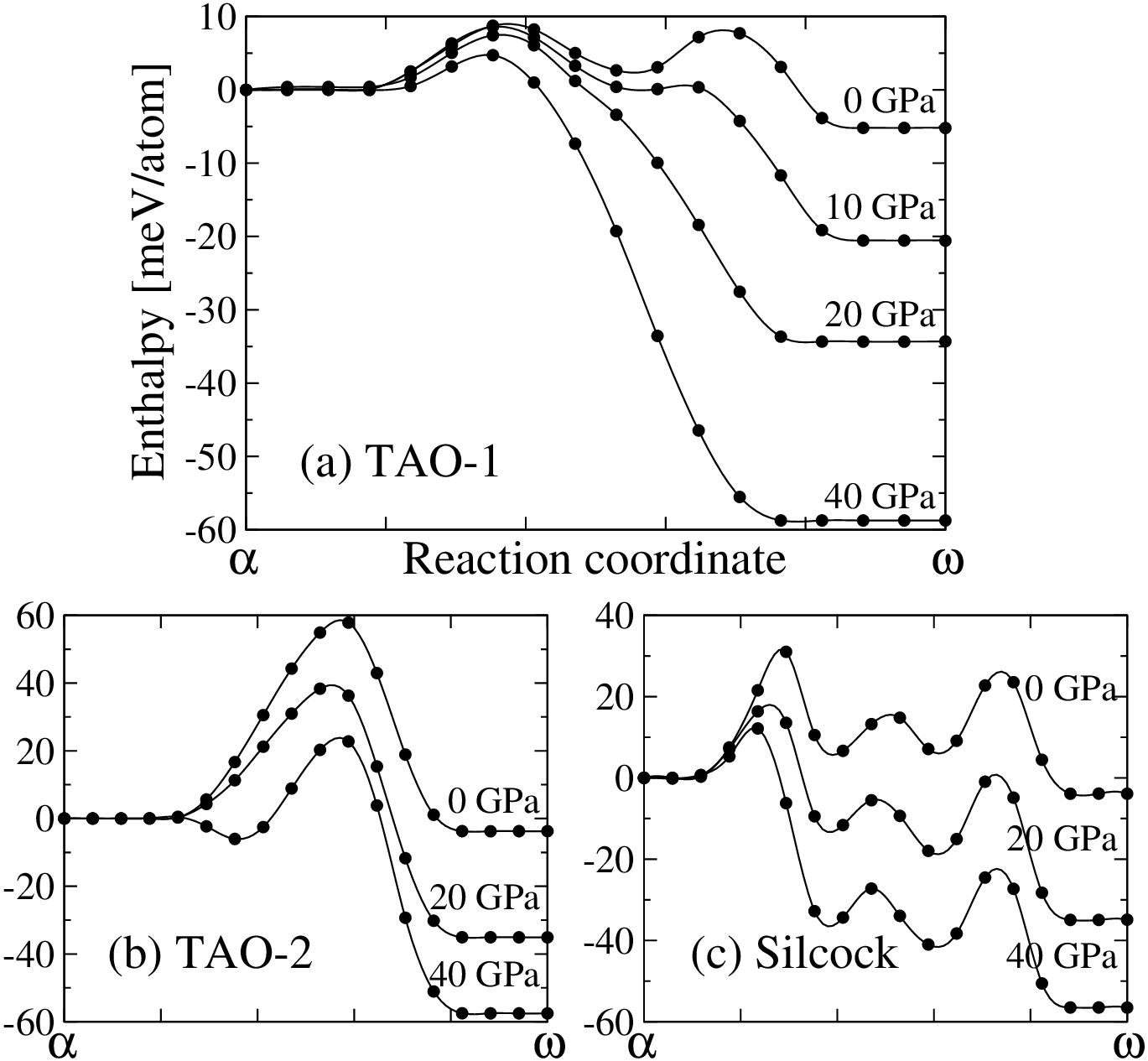}
\caption{\label{figEnBarrier} Enthalpy barrier vs. pressure for the
three lowest energy mechanisms, using \NEBVASP\ at 16 intermediate
states.  The \NEBVASP\ gives the $T=0$ energy along the homogeneous
\atow\ pathway for each mechanism.  The enthalpy barriers decrease
with increasing pressure.  (a) The TAO-1 mechanism gives the smallest
barrier by a factor of 4 at 0~GPa.  The TAO-2 mechanism (b) and
Silcock mechanism (c) have larger barriers at 0~GPa; an ordering which
continues even at 40~GPa.}
\end{figure}

Figure~\ref{figEnBarrier} shows the enthalpy along the
\atow\ transformation path for the three mechanisms as a function of
pressure.  Our calculation shows $\omega$ slightly lower in energy
than $\alpha$ at 0~GPa.  The crystal structure at 0~K has not been
determined experimentally; however, extrapolation of the
$\alpha$-$\omega$ phase boundary indicates $\omega$ as the ground
state~\cite{sikka}.  As pressure increases, the enthalpy of $\omega$
relative to $\alpha$ drops, and all three mechanisms decrease their
enthalpy barrier.  For the next four lowest landscape barrier
mechanisms in Figure~\ref{figEnComparison}, we find 0~GPa \NEBVASP\
barriers of 32, 37, 68, and 69~meV/atom.  The barrier of the TAO-1
mechanism is lowest, even up to 40~GPa.

We use classical nucleation theory to relate the homogeneous barriers
to the energy scales relevant for microscopic transformation
mechanics, given by the formation energy of a critical nucleus. If
the formation energy of the critical nuclei is of the same order as
the barrier, our mechanism prediction is not expected to change.
In classical nucleation theory, nuclei of new material grow only if
they are larger than a critical size where the nucleation energy is an
extrema. The nucleation energy consists of three terms: the
interfacial energy, the enthalpy difference, and the stress from the
volume difference.

In estimating the energy to create the critical nucleus, we use the
$\alpha$-Ti surface energy of 125~meV/\AA$^2$ as an upper bound for
the $\alpha$-$\omega$ interfacial energy.  The enthalpy difference at
the experimental shock transition pressure of 10~GPa is
$-20$~meV/atom, and the stress from the volume difference is
3.5~meV/atom.  We use the same critical nucleus size for each
mechanism in Table~\ref{table:mechEn} because they each have a plane
with no strain after the transformation, so the nuclei likely grow as
oblate spheroids to minimize stress.  Classical nucleation theory
estimates a critical nucleus radius of 240~\AA, (3.5 million atoms)
with a formation energy of 30~keV.  For that sized nucleus the
9~meV/atom barrier of TAO-1 becomes a homogeneous barrier of 28~keV.
In contrast, the competing TAO-2 and Silcock mechanisms have barriers
of 200 and 120~keV respectively; far above the formation energy and
the TAO-1 barrier.

Our systematic search, involving both tight-binding and \textit{ab
initio} calculations, combined with a nucleation analysis leads to the
direct TAO-1 mechanism as the preferred \atow\ mechanism in pure
titanium.  Our approach allows us to find the lowest barrier mechanism
for any martensitic phase transformation by generating and comparing
all relevant possible mechanisms.  Indeed this method provides an
invaluable starting point for investigating any martensitic
transformation.

\begin{acknowledgments}
\textbf{Acknowledgments.} 
We thank G.~T.~Gray III and C.~W.~Greeff for helpful discussions, and
M.~I.~Baskes and R.~B.~Schwarz for reviewing the manuscript. DRT
thanks Los Alamos National Laboratory for its hospitality and was
supported by a Fowler Fellowship at Ohio State University. This
research is supported by DOE grants DE-FG02-99ER45795 (OSU) and
W-7405-ENG-36 (LANL). Computational resources were provided by the
Ohio Supercomputing Center and NERSC.
\end{acknowledgments}

\end{document}